\documentclass[]{spie}  

\usepackage{amsmath,amsfonts,amssymb}
\usepackage{graphicx}
\usepackage[colorlinks=true, allcolors=blue]{hyperref}

\title{Prime-Cam: A first-light instrument for the CCAT-prime telescope}

\author[a]{E.~M. Vavagiakis}
\author[b]{Z.~Ahmed}
\author[c]{A.~Ali}
\author[d]{K.~Basu}
\author[e,f]{N.~Battaglia}
\author[d]{F.~Bertoldi}
\author[g]{R.~Bond}
\author[h]{R.~Bustos}
\author[i,j]{S.~C.~Chapman}
\author[k,b]{D.~Chung}
\author[m]{G.~Coppi}
\author[l]{N.~F.~Cothard}
\author[m]{S.~Dicker}
\author[a]{C.~J.~Duell}
\author[p]{S.~M.~Duff}
\author[d]{J.~Erler}
\author[n]{M.~Fich}
\author[o]{N.~Galitzki}
\author[a]{P.~A.~Gallardo}
\author[b]{S.~W.~Henderson}
\author[e]{T.~L.~Herter}
\author[p]{G.~Hilton}
\author[p]{J.~Hubmayr}
\author[b,k,q]{K.~D.~Irwin}
\author[a]{B.~J.~Koopman}
\author[r]{J.~McMahon}
\author[g]{N.~Murray}
\author[a]{M.~D.~Niemack}
\author[s]{T.~Nikola}
\author[g]{M.~Nolta}
\author[m]{J.~Orlowski-Scherer}
\author[e]{S.~C.~Parshley}
\author[e]{D.~A.~Riechers}
\author[e]{K.~Rossi}
\author[i]{D.~Scott}
\author[t]{C.~Sierra}
\author[o]{M.~Silva-Feaver}
\author[r]{S.~M.~Simon}
\author[e]{G.~J.~Stacey}
\author[a]{J.~R.~Stevens}
\author[p]{J.~N.~Ullom}
\author[p]{M.~R.~Vissers}
\author[p,u]{S.~Walker}
\author[v]{E.~J.~Wollack}
\author[m]{Z.~Xu}
\author[m]{N.~Zhu}

\affil[a]{Department of Physics, Cornell University, Ithaca, NY, USA 14853}
\affil[b]{Kavli Institute for Particle Astrophysics and Cosmology, SLAC National Accelerator Laboratory, Menlo Park, California, USA 94025}
\affil[c]{Department of Physics, University of California, Berkeley, Berkeley, CA, USA 94720}
\affil[d]{Argelander Institute for Astronomy, University of Bonn, D-53121 Bonn, Germany}
\affil[e]{Department of Astronomy, Cornell University, Ithaca, NY, USA 14853}
\affil[f]{Center for Computational Astrophysics, Simons Foundation, NYC, NY, USA 10010}
\affil[g]{Canadian Institute for Theoretical Astrophysics, University of Toronto, Toronto, ON, Canada}
\affil[h]{Facultad de Ingenier\'ia, Universidad Cat\'olica de la Sant\'isima Concepci\'on, Concepci\'on, Chile}
\affil[i]{Department of Physics and Astronomy, University of British Columbia, Vancouver, British Columbia, Canada}
\affil[j]{National Research Council, Herzberg Astronomy and Astrophysics, Victoria, British Columbia, Canada}
\affil[k]{Department of Physics, Stanford University, Stanford, CA, USA 94305}
\affil[l]{Department of Applied and Engineering Physics, Cornell University, Ithaca, NY, USA 14853}
\affil[m]{Department of Physics and Astronomy, University of Pennsylvania, Philadelphia, Pennsylvania, PA, USA 19104}
\affil[n]{Department of Physics and Astronomy, University of Waterloo, Waterloo, Ontario, Canada}
\affil[o]{Department of Physics, UCSD, La Jolla, CA, USA 92093}
\affil[p]{Quantum Sensors Group, National Institute of Standards and Technology, Boulder, CO, USA 80305}
\affil[q]{Kavli Institute for Particle Astrophysics and Cosmology, Stanford University, Stanford, CA, USA 94305}
\affil[r]{Department of Physics, University of Michigan, Ann Arbor, MI, USA 48109}
\affil[s]{Cornell Center for Astrophysics and Planetary Science, Cornell University, Ithaca, NY, USA 14853}
\affil[t]{Department of Applied Physics, University of Michigan, Ann Arbor, MI, USA 48109}
\affil[u]{Department of Astrophysical and Planetary Sciences, University of Colorado Boulder, Boulder, CO, USA 80309}
\affil[v]{NASA/Goddard Space Flight Center, Greenbelt, MD, USA 20771}

\authorinfo{Further author information: (Send correspondence to EMV: E-mail: ev66@cornell.edu, Telephone: 1 607 255-0474)}

\begin{document} 
\maketitle

\newpage 

\begin{abstract}
CCAT-prime will be a 6-meter aperture telescope operating from sub-mm to mm wavelengths, located at 5600 meters elevation on Cerro Chajnantor in the Atacama Desert in Chile. Its novel crossed-Dragone optical design will deliver a high throughput, wide field of view capable of illuminating much larger arrays of sub-mm and mm detectors than can existing telescopes. We present an overview of the motivation and design of Prime-Cam, a first-light instrument for CCAT-prime. Prime-Cam will house seven instrument modules in a 1.8 meter diameter cryostat, cooled by a dilution refrigerator. The optical elements will consist of silicon lenses, and the instrument modules can be individually optimized for particular science goals. The current design enables both broadband, dual-polarization measurements and narrow-band, Fabry-Perot spectroscopic imaging using multichroic transition-edge sensor (TES) bolometers operating between 190 and 450\,GHz. It also includes broadband kinetic induction detectors (KIDs) operating at 860\,GHz. This wide range of frequencies will allow excellent characterization and removal of galactic foregrounds, which will enable precision measurements of the sub-mm and mm sky. Prime-Cam will be used to constrain cosmology via the Sunyaev-Zeldovich effects, map the intensity of [CII] 158\,$\mu$m emission from the Epoch of Reionization, measure Cosmic Microwave Background polarization and foregrounds, and characterize the star formation history over a wide range of redshifts. More information about CCAT-prime can be found at \url{www.ccatobservatory.org}.
\end{abstract}


\section{INTRODUCTION}
\label{sec:intro}  

\begin{figure}[t!]
   \centering
   \includegraphics[width=1.0\textwidth]{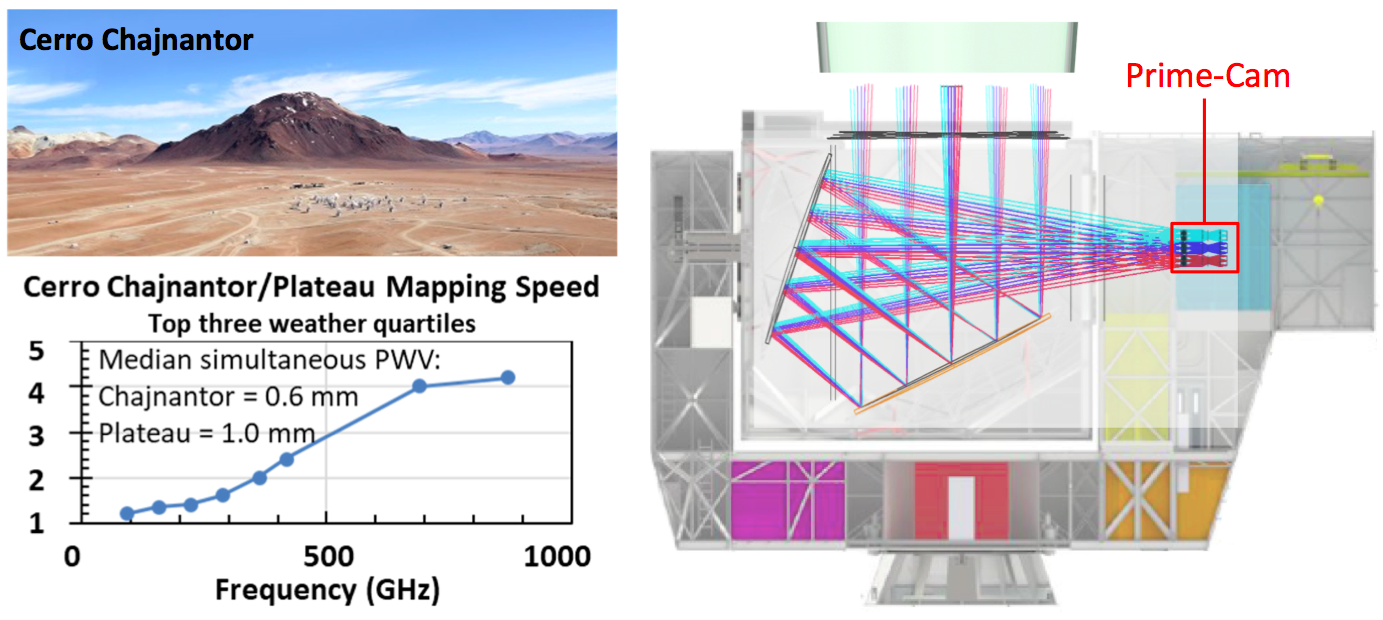}
   \centering
   \caption{{\it Left}: A photo of the CCAT-prime site, Cerro Chajnantor, and the improvement in sub-mm mapping speed expected at 5600\,m relative to that of the ALMA Plateau \cite{Radford:2016}. {\it Right}: A cross-section of the CCAT-prime telescope including the CCAT-prime optics, focusing onto the detector arrays of Prime-Cam, which sits in the telescope receiver cabin.}
  \label{fig:CCATpcam}
\end{figure}

The CCAT-prime telescope is a 6-m
aperture submillimeter (sub-mm) to millimeter (mm) wave telescope that is being built by an international consortium led by Cornell University, to be completed by
2021. CCAT-prime's extremely wide field of view (FoV) crossed-Dragone design will enable far faster mapping than existing facilities (Figure \ref{fig:CCATpcam}) \cite{niemack:2016,dragone}. The off-axis 6-m design achieves very low ($\sim1\%$) emissivity to take advantage of the Cerro Chajnantor site for sub-mm/mm wave astrophysics \cite{Bustos2014}. The high surface accuracy (half-wavefront error~$\leq$~10.7~$\mu$m) ensures excellent sub-mm sensitivity. The mirror shapes are modified to reduce coma, providing an FoV ranging from 2$^{\circ}$ in diameter at 860\,GHz (0.35\,mm) to 7.8$^{\circ}$  at 100\,GHz (3\,mm). The large FoV enables the illumination of approximately ten times more detectors than the current generation of millimeter telescopes.\cite{stacey:2018} 
The CCAT-prime telescope is described in Parshley~et~al. \cite{parshley:2018, parshley2:2018}.  This telescope concept was adopted by the Simons Observatory (SO) in 2017, and development of the telescope for SO has proceeded in conjunction with the CCAT-prime collaboration. Both telescopes are being built by Vertex Antennentechnik GmbH.  For Prime-Cam we have adopted cryogenic, optical, and detector readout designs from the Simons Observatory Large Aperture Telescope Receiver (LATR) \cite{galitzki:2018,zhu:2018,Scherer:2018,Coppi:2018,dicker:2018}.

As the first-light instrument for CCAT-prime, Prime-Cam will enable unique observations that address astrophysical questions ranging from the physics of star formation to Big Bang cosmology. Prime-Cam will simultaneously
cover five bands spanning 190 to 900\,GHz (1.6 to 0.33\,mm). 

The wavelength coverage, sensitivity,
spatial resolution, and large field of view (FoV) of Prime-Cam on CCAT-prime allow for a set
of wide-area surveys (between 5 and 15,000\,deg$^2$) to be conducted in order to address the following scientific goals:

\begin{enumerate}
\itemsep 0em
 
  \item Trace the formation and large-scale three-dimensional clustering of the first star-forming galaxies during the Epoch of Reionization through wide-field, broadband spectroscopy \cite{kovetz17};
  
  \item Constrain dark energy and feedback mechanisms by measuring the physical properties and distribution of galaxy clusters via the Sunyaev-Zeldovich (SZ) effects on the CMB \cite{mittal/etal:2017,Battaglia17};
  
  \item Enable more precise constraints on inflationary gravity waves and light relics by measuring polarized CMB foregrounds and Rayleigh scattering  \cite{abazajian/etal:2016,Alipour:2014dza};
  
\item Directly trace the evolution of dusty-obscured star formation in galaxies since the epoch of galaxy assembly, starting $>$ 10\,billion years ago \cite{riechers13c}.

\end{enumerate}

At first light, Prime-Cam will have three instrument modules optimized for wide-area surveys (5 -- 15,000\,deg$^2$) which are installed in a single cryostat. Two modules will make measurements in the 220, 270, 350 and 405\,GHz bands simultaneously using multichroic transition-edge sensor (TES) bolometers.   One of these modules is matched to the atmospheric windows for optimal detection of continuum radiation, while the other is equipped with an imaging Fabry-Perot interferometer (FPI) for spectral imaging observations.  A third module, enabled by the very low precipitable water vapor at the CCAT-prime site, will perform 860\,GHz continuum surveys utilizing kinetic inductor detectors (KIDs). \cite{Radford:2016,Bustos2014}

Table~\ref{tab:sciinst} shows the matching of science goals to the three first-light instrument modules and planned survey areas.  The required areas and integration times are based on simulations that include the wavelength coverage, sensitivity, and FoV of Prime-Cam.\cite{stacey:2018}

\begin{table}[h!]
\vspace{-5mm}

{\small
\hfill{}
\small
\begin{center}
\caption{Overview of how the science cases map onto the detector array types and preliminary survey areas. } 
\begin{tabular}{*8c}
\hline
Science cases$^a$ & Type$^b$ & Frequencies & Resolution & \multicolumn{2}{c}{Detectors} & \multicolumn{2}{c}{Survey Areas$^c$ (deg$^2$)} \\
 & & (GHz) & (arcsec) & Type & No. & Pilot & Full \\
\hline
SZ, CMB, SF & BB, Pol & 220, 270, 350, 405 & 53, 46, 39, 37  & TES & 8640 & 100 & 12,000 \\
EoR, SZ & Sp & 220, 270, 350, 405 & 53, 46, 39, 37 & TES & 4320 & 4 & 16 \\
SF & BB & 860 & 14 & KID & 18,216 & 100+4 & 12,000+16 \\ 
\hline
\label{tab:sciinst}
\end{tabular}
\vspace{-6mm}

\end{center}}
{\small
$^a$ Science cases: SZ, Clusters, SZ, \& Cosmology; CMB, CMB polarization; SF, Star Formation History; \\
 EoR, Intensity mapping of reionization. \\
$^b$ Array types: BB, broadband; Sp, spectroscopic with spectral resolution of 280; Pol, polarization sensitive. \\
$^c$ Preliminary areas for initial 400 hour and multi-year 4000 hour surveys for the three first-light modules. 860\,GHz will be used with both survey types.}
\end{table}

\section{THE PRIME-CAM RECEIVER}
\label{sec:receiver}

\begin{figure}[h!]
   \centering
   \includegraphics[width=1.0\textwidth]{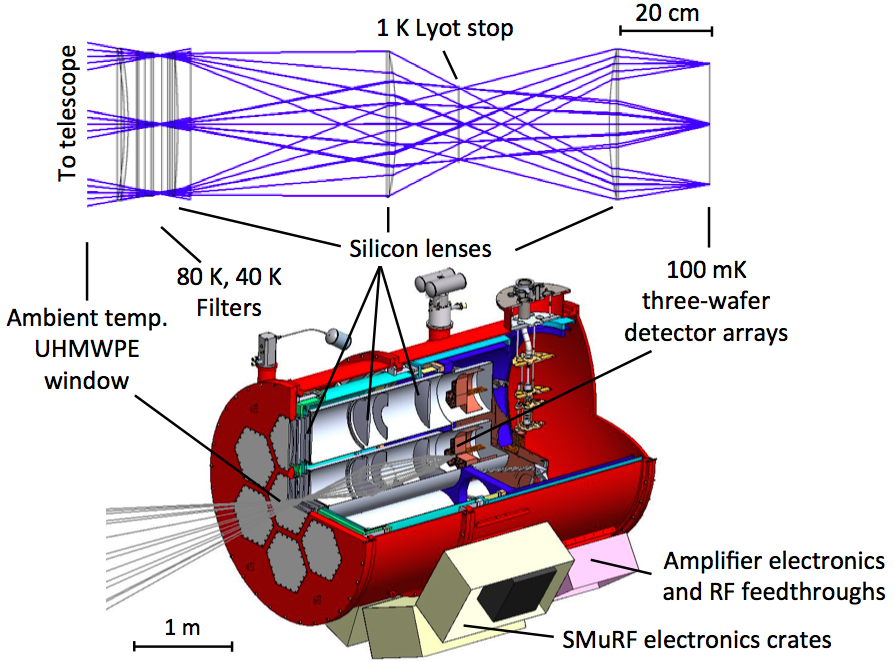}
   \vspace{1mm}
   \centering
   \caption{Summary of Prime-Cam's optical design. {\it Top:} Instrument module optics designs provide diffraction-limited image quality across a wide FoV. Cryogenic silicon lenses offer excellent optical performance at these wavelengths, and we have previously deployed silicon metamaterial AR coatings with $\textless$1\% reflection across an octave of bandwidth. \cite{datta/etal:2013,gallardo/etal:2017} {\it Bottom:} Preliminary design of the Prime-Cam instrument with three first-light instrument modules shown (two in cross-section). The refrigerators and room temperature SMuRF readout electronics crates are shown, as well as RF-sealed amplifier boxes that connect between the SMuRF electronics and cryostat feedthroughs. See Figure \ref{fig:xsec} for Prime-Cam instrument mechanical design details.}
    \label{fig:opticssum}
\end{figure}

The Prime-Cam design, which will have the ability to house seven instrument modules (aka ``optics tubes"), is shown in Figure \ref{fig:opticssum}. Section \ref{sec:optics} describes the three initial instrument modules designed for first-light observations. Section \ref{sec:dets} discusses the detector arrays, which are composed of polarization-sensitive TES bolometers and KIDs while Section \ref{sec:readout} describes the SLAC Microresonator Radio Frequency (SMuRF) detector readout electronics. Section \ref{sec:fpi} discusses the Fabry-Perot interferometer, called the Epoch of Reionization Spectrometer (EoR-Spec), used in one of the TES modules. Section \ref{sec:cryo} describes the Prime-Cam 1.8-meter diameter cryostat. Sections \ref{sec:ongoing} and \ref{sec:future} discuss ongoing work and future prospects.

\subsection{Optics}
\label{sec:optics}

\begin{figure}
   \centering
   \includegraphics[width=1.0\textwidth]{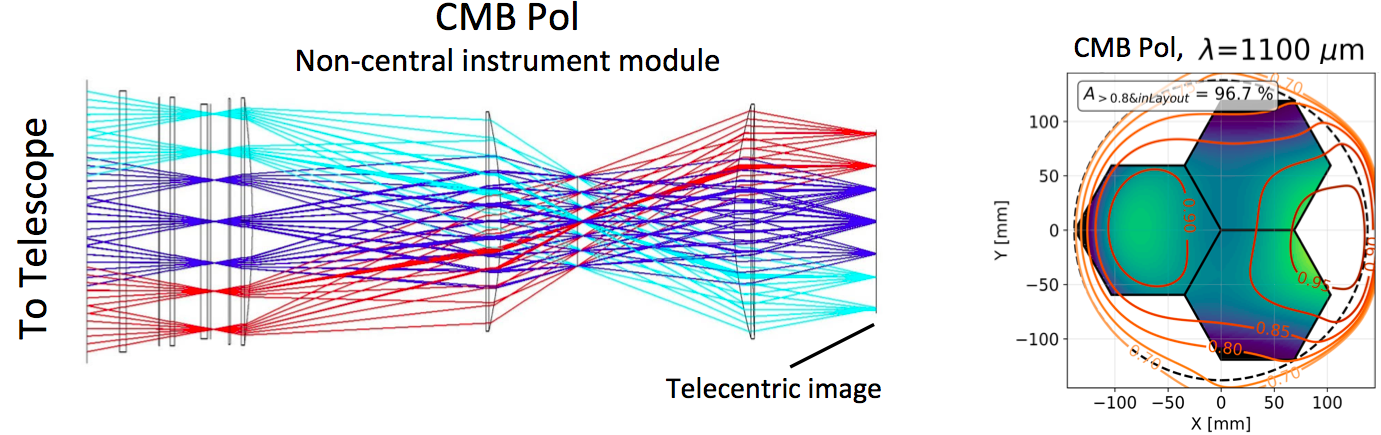}
   \vspace{5mm}
   \includegraphics[width=1.0\textwidth]{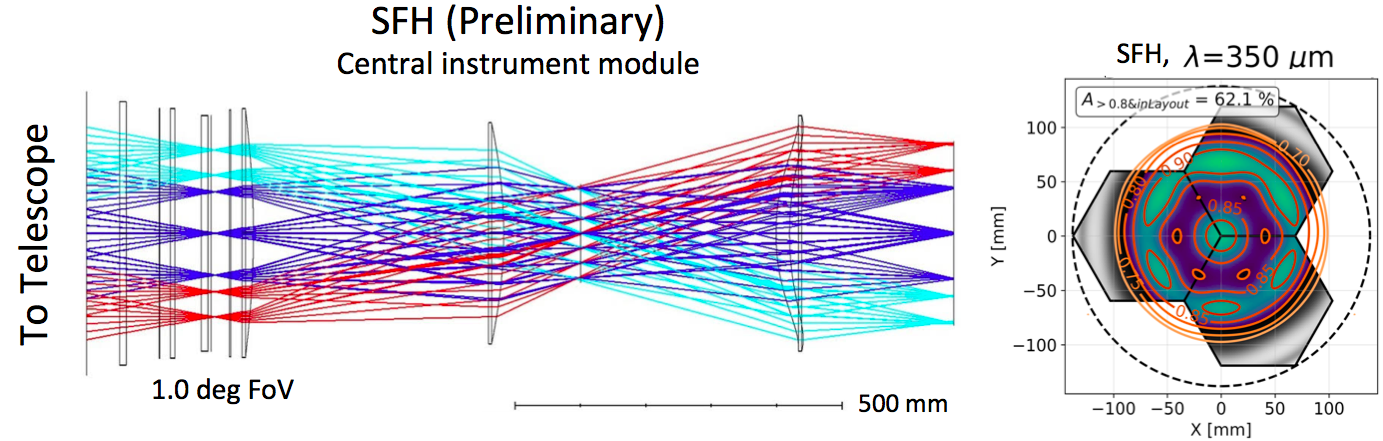}
   \vspace{5mm}
   \includegraphics[width=1.0\textwidth]{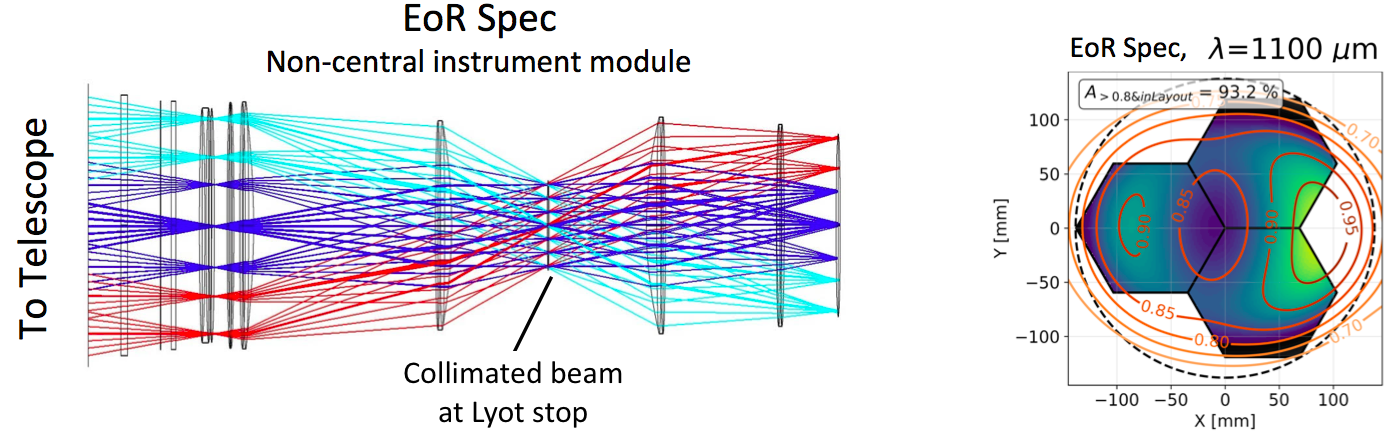}
   \centering
   \caption{Geometrical ray traces for each optics design for Prime-Cam's three initial instrument modules: the CMB polarization module (CMB Pol); the Epoch of Reionization spectrometer module (EoR-Spec), equipped with an FPI at the 1 K Lyot stop of the module and a four-lens design; and the star formation history (SFH) module, which has a 1.0$^{\circ}$ diameter FoV instead of the 1.3$^{\circ}$ diameter FoV of the other modules, providing better image quality at 350\,$\mu$m. Strehl ratios for each of the optics tubes are shown, with areas corresponding to the fraction of the area with Strehl ratio greater than 0.8 and within a tiled hexagonal pattern at the focal plane of the modules. One rotational degree of freedom, where the angle
of the tiled pattern at the focal plane is varied to allow maximum coverage, has the ability to be optimized in the future \cite{gallardo:2018}.}
\label{fig:raytraces}
\end{figure}

The modularity of Prime-Cam enables each optical path to be independently optimized for the science requirements of each instrument module (Figure \ref{fig:raytraces}). Each module within the Prime-Cam receiver contains a series of anti-reflection coated silicon optics, blocking filters, Lyot stop, and focal plane. The module optics are designed to provide diffraction-limited (or near diffraction-limited) image quality across a wide FoV. This instrument module, or ``optics tube," approach has been developed from the ACTPol model \cite{Thornton2016,Niemack2010} in collaboration with the Simons Observatory, which is building a receiver with 13 optics tubes that will all be used for broadband detection between 20 and 300\,GHz \cite{galitzki:2018,zhu:2018}.

Several different sizes of optics tubes were studied before converging on the approximately 0.4-m diameter optics tubes, each illuminating three 150-mm detector wafers, used in Prime-Cam.  The sizes studied ranged from approximately 0.2-m diameter optics tubes, each illuminating single 150-mm detector arrays, to 0.6-m optics tubes illuminating 7 detector arrays and  a single 2-m diameter optics tube illuminating many more detector arrays.  In addition to optical analyses, a sensitivity calculator that included estimates of the loss and emission of each component was used to compare all the configurations \cite{Hill:2018}. The conclusion of these analyses was that several configurations could provide similar mapping speeds, although, the image quality degrades as the optics tube diameter increases.  The 0.4-m diameter optics tube design provides a practical balance between good image quality, high throughput and mapping speed, modularity, manufacturability, and the ability to upgrade with new optics tubes. These qualities led the 0.4-m size to be selected for the primary instruments for both CCAT-prime and SO \cite{dicker:2018}.

Ultra-high-molecular-weight polyethylene (UHMWPE) vacuum windows and a series of approximately 6 infrared-blocking and low-pass filters minimize emission and block undesired radiation in the instrument \cite{tucker/ade:2006}. The telescope focus is transferred to the detector focal plane of each module by three refractive silicon lenses with metamaterial anti-reflection coatings (four lenses for EoR-Spec)\cite{datta/etal:2013}. Silicon is the preferred lens material at Prime-Cam's target wavelengths. High resistivity silicon has extremely low loss (tan\,$\delta$ $\sim$ $\times 10^{-5}$ at $T$ $<$ 40\,K, where $\delta$ is the loss angle), high thermal conductivity (ensuring lens temperature uniformity and limiting detector background loading), and a high index of refraction, $n$\,$\simeq$\,3.4\cite{datta/etal:2013}.

The baseline design for each instrument module is to target illuminating roughly 5.5\,m of the 6.0-m aperture telescope, which provides $f$/2.6 at the telescope focus. \cite{parshley:2018} The selected optics tube size provides an unobstructed 1.3\,deg diameter FoV and keeps the size of the entrance window manageable and within current fabrication capabilities.  

This full FoV can be focused onto three detector arrays at $f$/2.0 as shown in Figures \ref{fig:opticssum} and \ref{fig:raytraces}. This three lens instrument module optics design was optimized to achieve a telecentric, diffraction-limited image with minimal ellipticity for CMB polarization measurements across a 1.3$^{\circ}$ diameter FoV \cite{dicker:2018}. Similar designs have been optimized for use with an FPI, in which a larger optimization weight is given to collimating the light at the Lyot stop so as not to limit the finesse of the interferometer. This approach led to a preliminary design with four lenses instead of three (Figure \ref{fig:raytraces}) to improve the collimation. For the 860-GHz SFH module, a combination of ray tracing and physical optics calculations are being pursued to optimize the balance between resolution and sensitivity by adjusting the FoV and taking into account practical detector array and feedhorn geometry constraints. A three lens design with a 1.0$^{\circ}$ FoV (instead of 1.3$^{\circ}$), can provide significantly better (diffraction-limited) image quality at 350\,$\mu$m.  Reducing the FoV changes the effective pixel spacing at 350\,$\mu$m for the KID arrays, from 1.8 F-$\lambda$ to 1.4 F-$\lambda$, which improves the angular resolution at 350\,$\mu$m to near the diffraction-limited target of 14$^{\prime\prime}$.  While the three-lens design would suffice, optimization (including the consideration of a four-lens design) is ongoing for the SFH module to further improve the image quality and resolution.

\subsection{Detectors}
\label{sec:dets}

\begin{figure}[t!]
   \centering
   \includegraphics[width=1.0\textwidth]{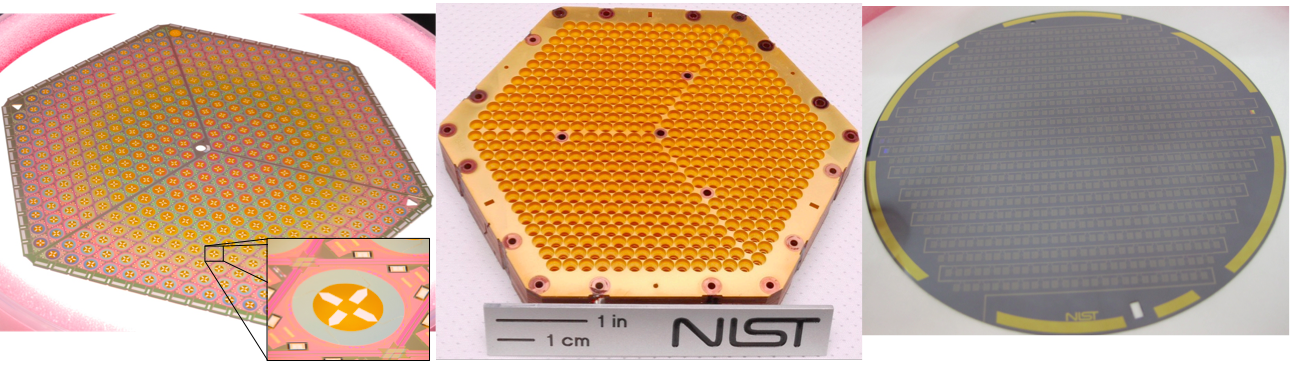}
   \centering
   \vspace{1mm}
   \caption{Prime-Cam's detector arrays will follow a heritage of AdvACT and BLAST arrays of TESes and KIDs. \textit{Left}: AdvACT polarization sensitive 150 mm 150/230\,GHz 2024 TES array, with a zoom in to a single polarization-sensitive pixel. \textit{Center}: AdvACT stacked silicon feedhorn array \cite{ward2016}. For Prime-Cam, we are baselining metal feedhorns. \textit{Right}: BLAST 100\,mm diameter 350\,$\mu$m Ti/TiN multilayer 938 KID array \cite{Galitzki2014,Lourie:2018}.}
   \label{fig:detarrays}
\end{figure}

Two focal planes with TESes and one with KIDs will be deployed as part of the first-light instrument. The first two focal planes will utilize silicon feedhorn arrays and multichroic (or multi-frequency) optical coupling technologies to measure four frequency bands simultaneously: one broadband, one spectroscopic \cite{datta/etal:2016,henderson/etal:2016}. These arrays will be based on TES detectors read out using superconducting quantum interference device (SQUID) multiplexers (Figure \ref{fig:detarrays}).\cite{dober/etal:2017}  

\begin{figure}[t!]
   \centering
   \includegraphics[width=1.0\textwidth]{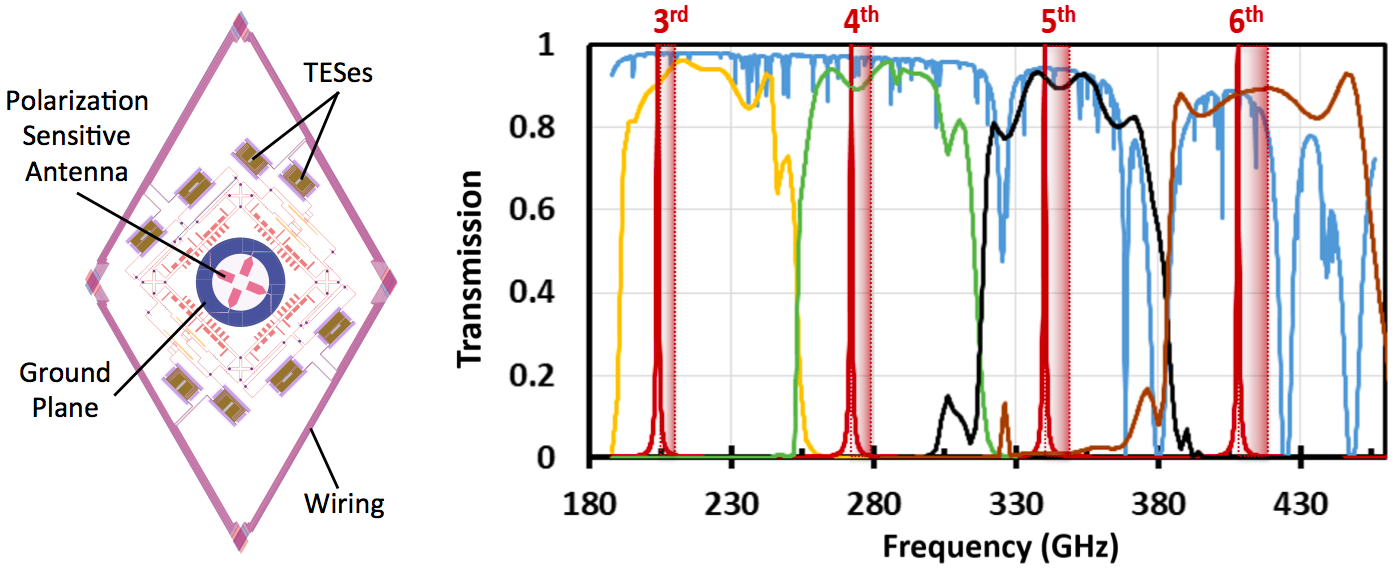}
   \centering
   \caption{\textit{Left}: Layout of a single quad-chroic pixel. The different TES designs are due to testing different termination schemes on each polarization. Most of the ground plane (blue) has been removed from the image for clarity. \textit{Right}: Four simulated multichroic passbands are shown (colors), along with the atmosphere transmission (blue) and the third through sixth order FPI fringes (red profiles), plus the instantaneous spectral coverage across the array (shaded red boxes). }
   \label{fig:detlayout}
\end{figure}

In both TES arrays, lithographically defined superconducting microwave components separate the two polarizations and four frequencies onto different microstrip lines. For the broadband detectors, the change in optical power for each polarization and frequency is converted to heat and measured with a superconducting TES bolometer. For the spectroscopic detectors a similar architecture will be used, but the two polarizations will be combined before measuring the power in each frequency channel with TES bolometers. The TES bolometers in the spectroscopic array must also have smaller saturation powers (1 -- 10\,pW) than the broadband detectors (20 -- 200\,pW). We have previously built devices with saturation powers that span these ranges \cite{koopman/etal:2017}.

The third focal plane type is optimized for single frequency measurements at 860 GHz. The shorter wavelength and resulting smaller pixel size leads to a different optimization. In particular, we expect to be able to fabricate 6,072 detectors for use at 860\,GHz on a single 15-cm diameter wafer, which is two to three times as many detectors as the TES arrays hold. The availability of these densely populated arrays, combined with recent progress in demonstrating photon-noise-limited performance with KIDs coupled to metal feedhorns,\cite{Zmuidzinas2012,hubmayr/etal:2015,dober/etal:2016,mckenney/etal:2018,austermann/etal:2018} motivate the selection of KIDs for the 860-GHz focal plane (Figure \ref{fig:detarrays}). Similar KIDs with metal feedhorns will be flown on the BLAST-TNG balloon project in late 2018, and they can be measured with the same readout electronics as the TES arrays (Figure \ref{fig:detarrays}) \cite{Lourie:2018}. Code to implement real-time monitoring of the detector arrays is being developed in conjunction with the Simons Observatory, and will enable rapid feedback on instrument characterization and operation of the system. 

The baseline designs for the EoR-Spec and CMB Pol modules have pixel apertures between 1.7 and 3.6 F-$\lambda$ at the central frequencies between 220 and 405\,GHz. These designs provide excellent 4-band sensitivity that is well matched for the EoR-Spec measurements. We are continuing to study the optimization for CMB polarization and galaxy cluster measurements. A smaller F-$\lambda$ spacing is not critical for these measurements; however, greater emphasis on improving the sensitivity of measurements above 300\,GHz is being considered in order to take advantage of the CCAT site and increase the complementarity with current and upcoming CMB observations. The CMB Pol module is expected to achieve similar sensitivity below 300\,GHz to what is described in Mittal et al. 2017.\cite{mittal/etal:2017} Deployment of multiple quad-chroic instrument modules or a module with $>$300\,GHz detector spacing near 1.5 F-$\lambda$ should enable the $>$300\,GHz sensitivities presented in Mittal et al. to be achieved.\cite{mittal/etal:2017} The preliminary SFH module design with a 1$^{\circ}$ FoV achieves 1.4 F-$\lambda$  spacing with 6072 detectors per wafer. We are continuing to study approaches to enable smaller F-$\lambda$ spacing at all wavelengths for future instrument modules, including adjusting the illuminated FoV with current detector array technologies or switching from quad-chroic to single-frequency detectors in each instrument module, which could use the same readout architecture. Another approach is to increase the number of detectors read out per focal plane, which could be accomplished using KIDs.

\subsection{Readout Electronics}
\label{sec:readout}

Prime-Cam will leverage recent improvements in readout
technology in order to achieve the high pixel counts of its planned TES and KID arrays. The latest generation of cold multiplexing technologies utilize improvements in superconducting microresonator development to greatly increase the number of sensors that can be read out per readout channel.

The Prime-Cam TES arrays will be read out using microwave SQUID ($\mu$SQUID) readout, in which each TES is inductively coupled through an RF-SQUID to its own microresonator at a unique microwave frequency (between $4$ and $8$~GHz)~\cite{Irwin2004,dober/etal:2017}. Thousands of TES-coupled resonators are then coupled to a shared feedline and read out simultaneously using standard RF readout techniques. KID arrays can use the same readout electronics as the TES arrays, but for KIDs, the KID itself is the resonator~\cite{Zmuidzinas2012}, which greatly simplifies the array design.

Prime-Cam will read out its arrays using the SLAC Microresonator Radio Frequency (SMuRF) electronics~\cite{Kernasovskiy2017}. The SMuRF system is being developed to enable the readout of the next generation of large low-temperature sensor arrays for X-ray and CMB detection. Each SMuRF ``blade'' aims to read 4000 TES or KID channels per coaxial pair over $4$~GHz of bandwidth. We plan for a more conservative readout of 2000 TESes or 3000 KIDs per SMuRF. The SMuRF electronics implements specialized FPGA algorithms for closed-loop tone tracking for each resonance, which increases the maximum achievable multiplexing factor of SMuRFs relative to other systems~\cite{vanRantwijk2016,Bourrion2011,Mazin2009,Duan2010,McHugh2012,Bourrion2012,Hattori2012,Yates2009}. System control will be accomplished using the Experimental Physics and Industrial Control System (EPICS) framework to facilitate communication between control PCs and the SMuRF electronics \cite{henderson:2018,Kernasovskiy2017}.

Data acquisition software to read out, characterize, and optimize the SQUID and detector readout systems is currently being developed, and will be used before instrument commissioning to perform lab testing of Prime-Cam's arrays. The CCAT-prime control system will provide an interface to allow time-stamping with scanning on the sky, allowing for the instrument array readout to be synchronized with the motion of the telescope. Adaptation of existing data reduction tools will enable the final conversion of the time domain data stream into on-sky maps while removing systematic noise sources (sky variation, bad pixels, correlated array noise, etc.).

\subsection{Epoch of Reionization Spectrometer}
\label{sec:fpi}

One of the two TES instrument modules will contain a spectrally/spatially multiplexing FPI, the Epoch of Reionization Spectrometer (EoR-Spec). This module is designed to measure the 158 $\mu$m [CII] line intensity from galaxies at redshifts between 9.3 and 3.3 (observed frequency from 190 to 450\,GHz). The spectrometer is optimized for surface brightness sensitivity with a beam well matched to the roughly 1 to few arcmin (0.5 to few Mpc) clustering scale of the galaxies producing the EoR signal. At 190 -- 450\,GHz, the best heterodyne receiver noise figures are 50 -- 75\,K (e.g. Wilson et al.\cite{Wilson2012}) and sky backgrounds are as low as 3\,K. Therefore the highest sensitivity is obtained with direct detection spectroscopy, where the fundamental noise limit is the statistical fluctuations in the photon arrival rates.

This high sensitivity must be properly calibrated and free of systematics over tens of square degrees, requiring robust suppression of systematics both in angle and frequency. Therefore, there is an advantage to a spectrometer that is highly multiplexed both spatially and spectrally. To properly resolve the intensity signal in redshift space, the spectrometer should have resolving power of 200 to 500, but it also needs to cover a factor of 2.4 in frequency space to map the [CII] line at $z$\,$\sim$\,9.3 to 3.3. 

\begin{figure}[t!]
   \centering
   \includegraphics[width=1.0\textwidth]{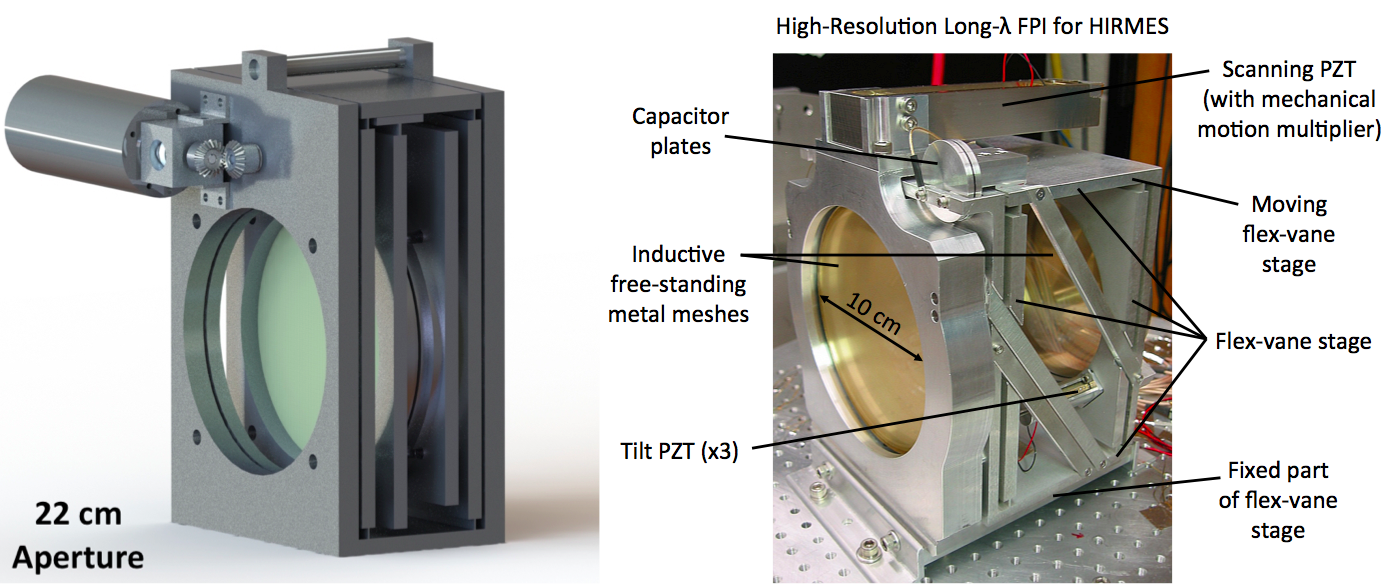}
   \centering
   \vspace{1mm}
   \caption{\textit{Left}: A prototype scaled 22-cm aperture cryomotor-driven Fabry-Perot Interferometer for Prime-Cam's spectroscopic instrument module, EoR-Spec. The FPI will be constructed from two anti-reflection coated silicon plates, one with a capacitive mesh, and one with an inductive mesh. Cryogenic silicon lenses and FPIs offer the best optical performance of any known material at these wavelengths. \textit{Right}: FPI for the HIgh Resolution Mid-infrarEd Spectrometer (HIRMES) for SOFIA.\cite{Douthit:2018,Kutyrev:2018} Compared to the FPI for Prime-Cam its dimensions in the cross-beam direction are about half, but along the beam are about the same. The basic components of the this FPI are very similar to the components for the Prime-Cam FPI and it can therefore be viewed as a prototype for the Prime-Cam FPI.}
   \label{fig:FPI}
\end{figure}

The EoR-Spec FPI maximizes sensitivity with high (\textgreater 70$\%$) cold transmission, delivering signal power to the thousands of background-limited detectors. The spectrometer derives from two technologies: a cryogenic scanning FPI with silicon substrate based (SSB) superconducting mirrors, and multichroic bolometers operating in the mm-wave regime. Inherent in the FPI/multichroic design is a high degree of cross-coupled spatial/spectral multiplexing: EoR-Spec employs 4320 TES bolometer sensors in 1080 spatial positions, each of which has four bolometers centered at 220, 270, 350 and 405\,GHz with 10 to 30$\%$ spectral bandwidth each.  Excepting atmospheric transmission gaps, the four bands are continuous. The FPI order of operation is selected such that four fringes of the FPI are detected, one in each band, simultaneously (Figure \ref{fig:detlayout}).  Due to conservation of etendue through our 20-cm optical pupil, off-axis detectors in the focal plane go through the FPI at an angle, so they see a bluer resonant wavelength than the on-axis pixels. The FPI therefore naturally spectrally multiplexes within each resonant order. The net result is that at one fixed FPI setting, 33\,GHz, or 13$\%$ of the 255\,GHz total bandwidth is sampled. With a resolving power (RP) of 300, this means that 38 resolution elements are sampled on the sky at any time, more blue-shifted at increasing distance from the optical axis. This provides the high degree of spatial and spectral cross-correlation that is necessary to recover the faint signal on large scales. 

We will obtain our final spectral-spatial data cube by fixing the FPI at a given central frequency, rapidly scanning the sky spatially for a few minutes, then stepping across the entire frequency range one step at a time until the entire spectrum is obtained to the required depth.  To sample the full 262\,GHz bandwidth of our science requires around 10 fixed settings of the FPI. To fully sample the lower redshifts ($z$\,$\sim$\,3.3 to 5.2) will only take six spectral settings.

Our previous generation FPIs are based on free-standing metal mesh screens \cite{Poglitsch1991,Stacey1991,Latvakoski1997,Latvakoski1999, Bradford:2002, Oberst:2009,Douthit:2018,Kutyrev:2018}, where the screen geometry results in a finesse that is a strong function of frequency ($F$ $\propto \nu^{-2.5\:\mathrm{to}\:3}$). Due to transmission losses with too high a finesse, and spectral purity and resolving power losses with too low a finesse, this would limit efficiency over the octave bandwidth of operation we require. To avoid these losses, we will use the silicon substrate based (SSB) mirror designs under development at Cornell \cite{gallardo/etal:2017,cothard:2018}. The nanofabrication techniques inherent in SSB mirrors enable us to tune the reflective geometry and finesse. We have investigated geometries for our reflectors including pure inductive (screen-like), pure capacitive (isolated metal islands) and mixed combinations of the two for a wide variety of metal fill to open silicon values.  Our best result occurs for a plane-parallel silicon etalon based on one reflector with an inductive grid (reflectivity rises with wavelength), and the other with a capacitive grid (reflectivity falls with wavelength). Modeling results of a complete SSB mirror (AR coat on one side, metalized on the other) already show good performance over the octave bandwidth of astrophysical interest -- finesse 30 to 60, and transmission 65 $\mathrm{to}$ 98$\%$.  We expect to achieve higher performance with further optimization. 

Our cryogenic FPI mechanism (Figure \ref{fig:FPI}) is based on the deformable parallelogram translation stage we have used in previous generations of FPI systems \cite{Poglitsch1991,Latvakoski1997,Bradford:2002,Oberst:2009,Douthit:2018,Kutyrev:2018}. The parallelogram translation stage consists of two vertical ``flex vanes'' attached via thin wafer flexures to the horizontal top and bottom of the structure. The mirrors in a parallelogram-based translation stage can lose parallelism when the stage is driven if the lengths of the opposing sides of the parallelogram are not precisely equal. To ensure adequate precision, our parallelogram will be wire electron discharge machined from a single block of thermally annealed aluminum. The bottom of the parallelogram is attached to the outer walls of the FPI device. The left mirror is attached to the scanning stage, and the right plate is fixed on the structure frame-plate on stand-offs that penetrate the right flex vane. The frame-plate-mounted mirror can be adjusted in tilt with three fine thread (80 tpi) screws. Translation of the stage is effected by tuning the lead screw whose nut is attached to the top of the translation stage. The step size is designed to over-sample an FPI resolution element by a factor of 5. This requirement is exceeded by direct drive of the fine thread lead screw with a cryogenic stepper motor. Temperature monitoring and instrument control software will be implemented based on code used for previous experiments operating at 100\,mK, such as AdvACT and ZEUS.\cite{zeus2014} 

The FPI will be installed in a 20-cm pupil of the optical system. This image of the primary mirror is small enough that off-axis pixels will be shifted by several resolution elements in frequency (to the blue) to widen our instantaneous bandwidth, but large enough that resolving powers up to 1000 (far greater than our $R$~$\sim$200~to~500 requirement) can be delivered over the entire FoV.  The FPI can take in both polarizations and more than one spatial mode of the radiation field so that our FPI approach has throughput advantages that can more than double the per-pixel sensitivity (or more than quadruple the mapping speed/pixel) over waveguide based spectrally multiplexed or ``on-chip" resonator approaches {\cite{Crites:2014}. Our present design takes in a single spatial mode while maintaining the mapping speed equivalent to a resonant spectrometer with 12,000 bolometers. Our single module will deliver the EoR intensity mapping science. The FPI centered on 220 GHz will find additional use in measuring the ``zero crossing" of the thermal Sunyaev-Zeldovich (SZ) effect spectrum towards several massive galaxy clusters. This type of measurement will open a new frontier in the SZ spectral separation and extraction of cluster parameters like temperature and peculiar velocities, which will be complementary to our broadband measurements using multiple frequency modules.

\begin{figure}[h!]
   \centering
   \includegraphics[width=1.0\textwidth]{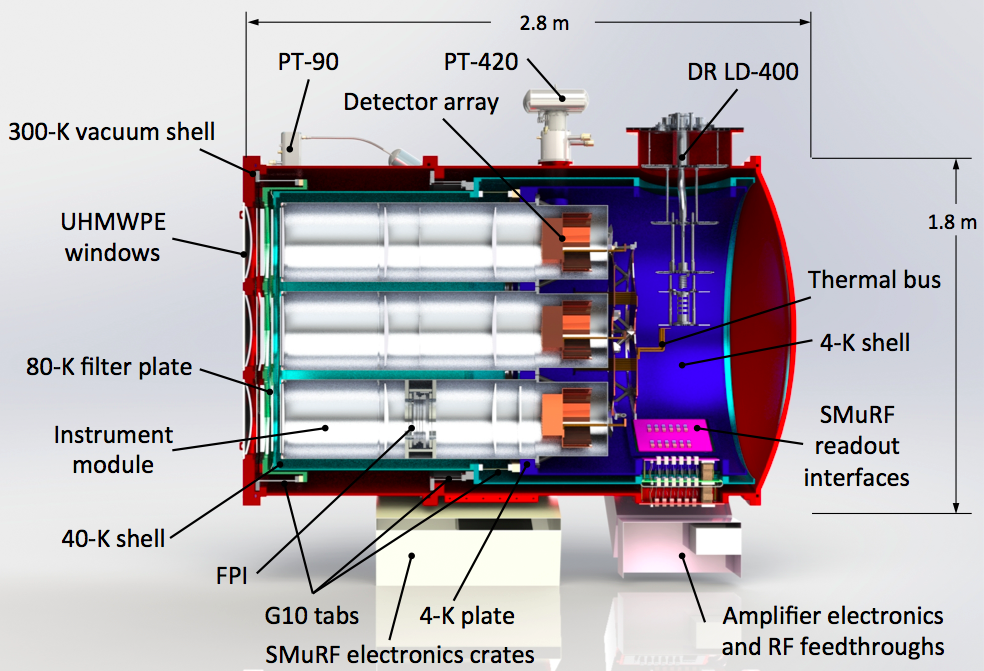}
   \vspace{1mm}
   \centering
   \caption{A rendered cross section of the Prime-Cam cryogenic receiver design. The two-section 6061-T6 Aluminum vacuum shell is shown in red. An 80-K filter plate and ring is shown in green, followed by a 40-K plate and shell in light blue, a 4-K plate and shell in dark blue, and thermal buses in copper. The temperature stages are supported by a series of G10 tabs. The DR, two pulse tubes, three instrument modules, and detector arrays of the initial three tube deployment are shown. SMuRF electronics crates and readout interfaces are also present.}
   \label{fig:xsec}
\end{figure}

\subsection{Cryogenics}
\label{sec:cryo}

The Prime-Cam cryogenic receiver will consist of a two-section 6061-T6 aluminum vacuum shell, 1.8\,m in diameter, and 2.8\,m in length (Figure \ref{fig:xsec}). The seven optics tube receiver design has evolved from the larger thirteen tube cryogenic design for the Simons Observatory LATR \cite{galitzki:2018,zhu:2018}. Hexagonal UHMWPE vacuum windows on the front plate will begin the optical chain of the instrument. Within the vacuum shell, several cooling stages provide thermal isolation for the optics tubes and detector arrays. The temperature and stability requirements for the cryostat interior are derived from the requirements for the SO LATR \cite{Scherer:2018,Coppi:2018}. In addition to the temperature requirements, the cryostat must be mechanically stable under the roughly 0.1-g load encountered during execution of the nominal science scan modes of the telescope. 

 80-K and 40-K temperature stages will hold optical filters. A short 80-K shield will be located at the front of the receiver. A 40-K shield will surround the interior of the receiver. A series of G10 tabs will support the assembly of thermal shields and provide thermal isolation as well as resilience against mechanical shocks and vibration. Inside the 40-K shield, the three optics tubes for initial deployment will be mounted on the 4-K temperature stage of the cryostat. A 4-K shield will surround the back end of the optics tubes. Cryomech PT-90 and PT-420 pulse tubes, each providing 90\,W of cooling power at 80\,K (PT-90), 55\,W at 40\,K (PT-420), and 2\,W at 4\,K (PT-420), along with a Bluefors LD-400 dilution refrigerator (DR) providing 400\,$\mu$W of cooling power at 100\,mK, will be used to cool the receiver. The PT-90s will be located at the front of the receiver, and will cool the 80-K plate and shield. The PT-420s will be located towards the rear of the cryostat, and will cool the 40-K shield and 4-K plate. The DR will be located at the rear of the cryostat, and will cool the detector arrays to 100\,mK and the final optical lenses to 1 K through thermal buses. Around the exterior of the vacuum shell will lie supports for the room temperature SMuRF readout electronics crates. Three SMuRF readout towers will be mounted on the vacuum shell and extend to within the 4-K shell, thermally insulated with multi-layer insulation, allowing for up to 30 SMuRFs able to read out about 60,000 TESes or 120,000 KIDs. The fully loaded receiver, including seven optics tubes and all readout components, will weigh about 2.8 metric tons. 

\subsection{Ongoing Work}
\label{sec:ongoing}

Prime-Cam's cryogenic design is nearing completion, with finite element analysis studies being done to determine the final design of the vacuum vessel (Figure \ref{fig:fea}). Simulations with ANSYS Maxwell are being performed in order to design the magnetic shielding for Prime-Cam's instrument modules (Figure \ref{fig:mag}, Table \ref{table2}). 

\begin{table}

{\small
\hfill{}
\small
\begin{center}
\begin{tabular}{*5c}
\hline
Tube & No Overall Shield  & No Overall Shield & Overall Shield & Overall Shield  \\
& (Axial) & (Transverse) & (Axial) & (Transverse) \\
\hline
Center Tube & 380 & 220  & 6670  & 8430 \\
Outer Tube & 160 & 220 & 2060 & 8620 \\
\hline

\end{tabular}
\vspace{2mm}

\end{center}}
\caption{Simulated magnetic shielding factors at the locations of detector arrays for two Amumetal 4K (A4K) shielding geometries and axial or transverse applied magnetic fields. The inclusion of a room temperature overall shield in addition to the individual optics tube shields greatly increases shielding factors, but due to design and cost constraints we will not be pursuing this configuration. The individual optics tube shields in combination with local device shields within the optics tubes provide sufficient shielding for our application.}
\label{table2}
\end{table}

\begin{figure}[h!]
   \centering
   \vspace{3mm}
   \includegraphics[width=0.60\textwidth]{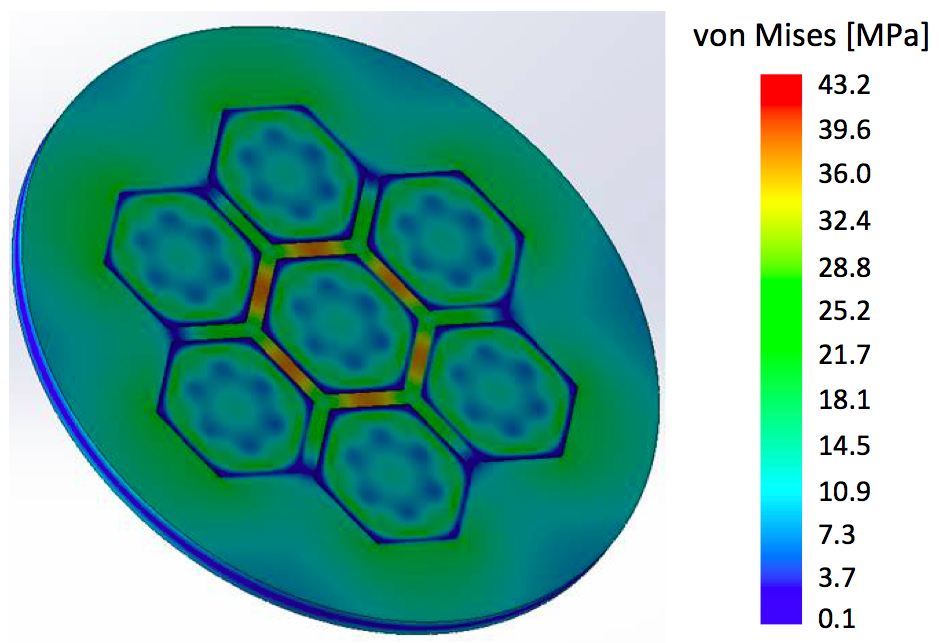}
   \vspace{2mm}
   \centering
   \caption{Finite element analysis results for Prime-Cam's vacuum window plate. The highest stresses of 43.2 MPa are located at supports surround the central window, below the 6061-T6 Al yield stress of 275 MPa. Analyzed with a factor of safety of 1.5 and resulting in a positive margin of safety of 3.2, the plate is well within our design parameters.}
    \label{fig:fea}
   \vspace{6mm}
\end{figure}

\begin{figure}[h!]
   \centering
   \includegraphics[width=1.0\textwidth]{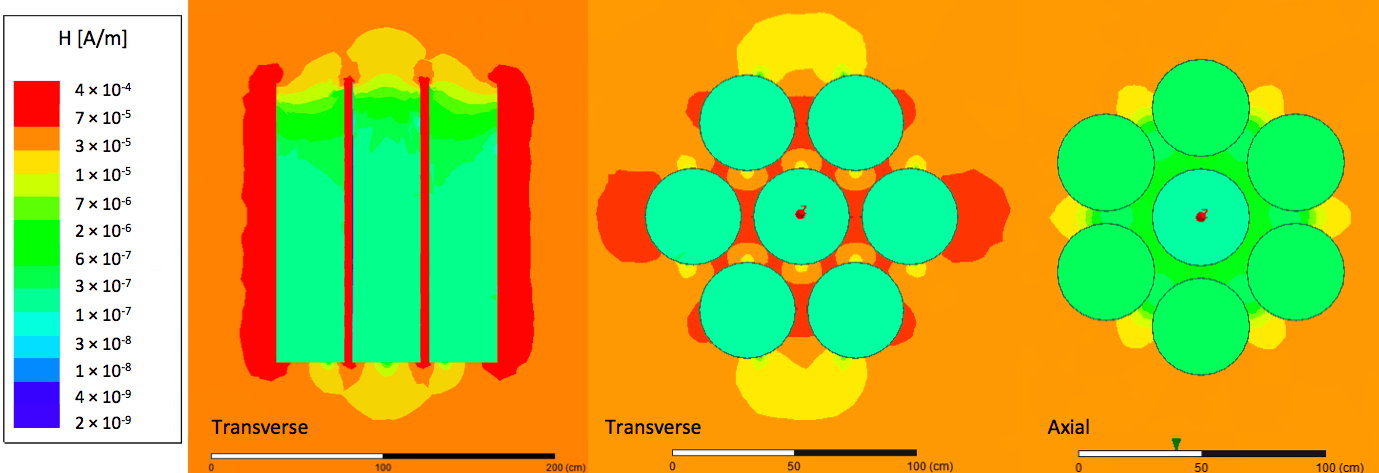}
     \vspace{1mm}
   \centering
   \caption{Magnetic shielding factor estimates for a Prime-Cam geometry seven optics tube design employing Amumetal 4K (A4K).}
   \label{fig:mag}
   \vspace{1mm}
\end{figure}

\subsection{Future Directions}
\label{sec:future}

Prime-Cam's modular receiver design offers flexibility for future upgrades. Adding more instrument modules can provide new measurement capabilities or reduce integration time by improving sensitivity.  Sub-millimeter sensitivity could be improved by adding single-frequency or dichroic instrument modules dedicated to the sub-millimeter bands, with more detectors than the quad-chroic CMB Pol arrays. Given the challenges of reading out large arrays of TESes, KID arrays may eventually be needed to realize the full sub-millimeter capabilities of this instrument. Observations at new frequencies, including 90\,GHz, 150\,GHz, 750\,GHz, and 1.5\,THz would be valuable for both complementing the current science goals\cite{Erler2018,mittal/etal:2017}  and enabling new goals.

Beyond broadband surveys, additional spectrometers could be deployed, with the possibility of including up to seven spectroscopic modules to fill the Prime-Cam receiver. Such an approach, or one including on-chip spectrometers, could dramatically improve future measurements of [CII] from the Epoch of Reionization or other spectroscopic measurements \cite{SuperSpec:2018}. Beyond Prime-Cam, CMB-S4 is the natural extension of the approach of deploying more detectors and instrument modules, and would benefit from the development of an even larger diameter cryostat for CCAT-prime with the capability to hold up to 19 Prime-Cam or SO-style optics tubes \cite{CMBS4TechnologyBook}.




\section{CONCLUSION}
\label{sec:conclusion}

Prime-Cam, the first-light instrument for CCAT-prime, will observe over the spectral range of 190 -- 900 GHz with sufficient sensitivity, spatial resolution and field of view to make strides towards understanding the history of the first star-forming galaxies, measuring polarized CMB foregrounds and Rayleigh scattering, tracing the evolution of dust-obscured star formation in young galaxies, and constraining dark energy and feedback mechanisms via the Sunyaev-Zeldovich effects on the CMB. 

Prime-Cam's detector arrays of TESes will simultaneously image in four bands (220, 270, 350 and 405~GHz), and spectroscopic capabilities will be enabled by employing a 1$^{\circ}$ FoV imaging Fabry-Perot interferometer (FPI). The Epoch of Reionization Spectrometer will be based on etalons optimized for broadband spectroscopy that multiplex in the four telluric windows. Prime-Cam's highest frequency 860-GHz camera will utilize KIDs, with observations enabled by the very low precipitable water vapor at the CCAT-prime site. Prime-Cam's wide frequency coverage and spectroscopic wide-field, broadband submillimeter-wave spectroscopy will allow for excellent measurements and characterization of foregrounds.  

The upgradable Prime-Cam cryogenic receiver is cooled by pulse tubes and a dilution refrigerator and is capable of housing up to seven instrument modules. Three modules will be deployed as part of this first-light instrument. The first-light observation target is early 2021.

\acknowledgments 
 
EMV was supported by the NSF GRFP under Grant No. DGE-1650441. CCAT-prime funding has been provided by Cornell University, the Fred M. Young Jr. Charitable Fund, the German Research Foundation (DFG) through grant number INST 216/733-1 FUGG,  the Univ.~of Cologne, the Univ.~of Bonn, and the Canadian Atacama Telescope Consortium. MDN acknowledges support from NSF award AST-1454881.

\bibliography{SPIE2018} 
\bibliographystyle{spiebib} 

\end{document}